\documentclass[%
 reprint,
 amsmath,amssymb,
 aps,
prb,
]{revtex4-1}

\usepackage[T1]{fontenc}
\usepackage[utf8]{inputenc}
\usepackage[english]{babel}
\usepackage{soul}
\usepackage[usenames,dvipsnames]{xcolor}
\usepackage[version=4]{mhchem}
\usepackage{graphicx}% Include figure files
\usepackage{dcolumn}% Align table columns on decimal point
\usepackage{bm}% bold math
\usepackage{hyperref}
\hypersetup{colorlinks=true, linkcolor=blue}

\begin{document}

\title{Surface and interface effects in oxygen deficient \texorpdfstring{\ce{SrMnO3}}{SrMnO3} thin films grown on \texorpdfstring{\ce{SrTiO3}}{SrTiO3}}

\author{Moloud Kaviani}
\email{moloud.kaviani@dcb.unibe.ch}
\affiliation{Department of Chemistry, Biochemistry and Pharmaceutical Sciences, University of Bern, Freiestrasse 3, CH-3012 Bern, Switzerland}

\author{Ulrich Aschauer} 
\email{ulrich.aschauer@dcb.unibe.ch}
\affiliation{Department of Chemistry, Biochemistry and Pharmaceutical Sciences, University of Bern, Freiestrasse 3, CH-3012 Bern, Switzerland}

\begin{abstract}
Complex oxide functionality, such as ferroelectricity, magnetism or superconductivity is often achieved in epitaxial thin-film geometries. Oxygen vacancies tend to be the dominant type of defect in these materials but a fundamental understanding of their stability and electronic structure has so far mostly been established in the bulk or strained bulk, neglecting interfaces and surfaces present in a thin-film geometry. We investigate here, via density functional theory calculations, oxygen vacancies in the model system of a \ce{SrMnO3} (SMO) thin film grown on a \ce{SrTiO3} (STO) (001) substrate. Structural and electronic differences compared to bulk SMO result mainly from undercoordination at the film surface. The changed crystal field leads to a depletion of subsurface valence-band states and transfer of this charge to surface Mn atoms, both of which strongly affect the defect chemistry in the film. The result is a strong preference of oxygen vacancies in the surface region compared to deeper layers. Finally, for metastable oxygen vacancies in the substrate, we predict a spatial separation of the defect from its excess charge, the latter being accommodated in the film but close to the substrate boundary. These results show that surface and interface effects lead to significant differences in stability and electronic structure of oxygen vacancies in thin-film geometries compared to the (strained) bulk.
\end{abstract}

\maketitle

%%%%%%%%%%%%%%%%%%%%%%%%%%%%%%%%%%%%%%%%%%%%%%%%%%%%%%%%%%%%%%%%%%%%%%%
\section{Introduction}
%%%%%%%%%%%%%%%%%%%%%%%%%%%%%%%%%%%%%%%%%%%%%%%%%%%%%%%%%%%%%%%%%%%%%%%

Transition-metal perovskite oxides represent an extremely versatile class of materials that can host a large range of functional properties such as ferroelectricity, magnetism or superconductivity~\cite{PhysRevX.3.021002, PhysRevMaterials.2.084414}. The emergence of these properties can often be tuned by bi-axial strain, imposed for example by lattice matching during coherent epitaxial growth on a substrate with different lattice parameter~\cite{Mannhart1607}. Compared to bulk perovskite oxides, fundamental changes in properties occur in these thin films. Ferroelectricity and magnetism can, for example, be enhanced~\cite{Wang:2003ca} or even introduced in the thin-film material~\cite{Choi:2012kp, PhysRevB.75.144402}. Moreover, interfaces between the substrate and the film or between different layers of a heterostructure have emerged as an avenue to generate rich and novel electronic phases~\cite{Mannhart1607}.

Depending on the synthesis conditions, complex oxides typically contain point defects that can strongly affect conductive, ferroelectric or magnetic properties useful for applications in electronics. While often detrimental to functional properties, defects were also shown to induce novel functionalities in specific cases~\cite{Sharma2017Designing, Becher2015}. Oxygen vacancies (\ce{V_O}) are particularly abundant in perovskite oxides under typical synthesis conditions. In their neutral charge state (\ce{V_O^{..}} in Kröger-Vink notation~\cite{Kroger:1956bl}) these defects lead to changes in oxidation state and local distortions that affect the ferroelectric and magnetic properties~\cite{Choi2013, Goodenough_2004, TAGUCHI1979221}. Point-defect engineering could thus be a route to tailor properties for a given application. Our understanding of point defects and their formation energetics and electronic structure is, however, currently mostly limited to idealized bulk or strained bulk systems, neglecting the effect of the substrate-film interface and the film surface, except for select cases like the LAO/STO interface~\cite{Yu2014discovery}. Therefore, realistic models containing both surfaces and hetero-interface with the substrate are crucially needed to accurately assess defect-induced phenomena in thin-film systems.

In the present work, we use the model system of a \ce{SrMnO3} (SMO) thin film grown on a typical (001) \ce{TiO2}-terminated \ce{SrTiO3} (STO) substrate to study the formation and electronic structure of \ce{V_O} in a thin-film geometry. This model system has recently been experimentally realized, showing excess charge accommodation in the film but close to the interface~\cite{Wang2020}. The thermodynamic ground state of SMO is a hexagonal phase~\cite{Negas1970}, but it is synthesizable in the perovskite structure when grown on a perovskite structured substrate~\cite{Becher2015, Maurel:2015hx}, possibly adopting ordered Brownmillerite phases at high oxygen deficiency~\cite{Kobayashi:2013gz}. \ce{V_O} were previously studied in both materials separately. In SMO and related manganites such as \ce{CaMnO3} and \ce{BaMnO3}, the three 3\textit{d} electrons in \ce{Mn^4+} fully occupy the $t_{2g}$ orbitals. Upon \ce{V_O} creation, the $e_g$ orbitals of \ce{Mn} adjacent to the defect are stabilized and accommodate the two excess electrons, resulting in a reduction from \ce{Mn^4+} to \ce{Mn^3+}~\cite{PhysRevB.88.054111, Becher2015, marthinsen_faber_aschauer_spaldin_selbach_2016}. It was also shown that magnetic order and polar distortions can affect the formation energy of \ce{V_O} in these materials~\cite{marthinsen_faber_aschauer_spaldin_selbach_2016, ricca2019self}. In STO, a large variability in the experimental and theoretical literature reveals that states with excess charge accommodation in the $t_{2g}$ conduction band or in shallow $e_g$ defect states are close in energy, results depending also on the crystal structure (cubic or tetragonal) and the size of the simulation cell~\cite{PhysRevResearch.2.023313}.

For the thin-film geometry, our DFT$+U$ calculations show that crystal-field changes of the under-coordinated surface atoms lead to a charge transfer towards the surface and hence surface \ce{Mn^3+} species. This asymmetric structure leads to an electric field in this nominally non-polar interface, which strongly affects the defect chemistry. We find that vacancies more easily form at the surface and that the formation energy increases in a near linear fashion with increasing distance from the surface. For oxygen vacancies in the STO substrate, we predict separation of the defect and the charge, the latter residing on Mn atoms at the interface, which leads to a marked reduction in formation energy compared to bulk STO. For our model system, the formation of oxygen vacancies is, therefore, greatly different compared to either of the bulk materials.

%%%%%%%%%%%%%%%%%%%%%%%%%%%%%%%%%%%%%%%%%%%%%%%%%%%%%%%%%%%%%%%%%%%%%%%
\section{Methods}
%%%%%%%%%%%%%%%%%%%%%%%%%%%%%%%%%%%%%%%%%%%%%%%%%%%%%%%%%%%%%%%%%%%%%%%

DFT calculations were performed using \textsc{Quantum} ESPRESSO~\cite{Giannozzi_2009, Giannozzi:2017io} at the PBE+$U$ level of theory~\cite{PhysRevLett.77.3865, PhysRevB.44.943} with Hubbard $U$ calculated self-consistently~\cite{PhysRevB.71.035105, PhysRevB.79.125124, PhysRevB.98.085127} as 4.26 and 4.48 eV for the Mn and Ti 3d orbitals respectively~\cite{ricca2019self, PhysRevResearch.2.023313}. All atoms are represented by ultrasoft pseudopotentials~\cite{PhysRevB.41.7892} with Sr(4s, 4p, 5s), Mn(3p, 4s, 3d), O(2s, 2p) and Ti(3s,3p, 3d, 4s) valence electrons. The cutoff for the plane-wave basis was 70 Ry for the kinetic energy combined with 840 Ry for the augmented density.

We model a SMO thin film, grown epitaxially on a (001) STO substrate. Unstrained perovskite SMO has a paraelectric \textit{Pnma} structure with a G-type antiferromagnetic order~\cite{PhysRevB.64.134412}. The structure is close to the ideal cubic structure with small octahedral tilts and rotations found computationally but not yet observed experimentally~\cite{PhysRevLett.104.207204}. The STO substrate undergoes, around 105K~\cite{COWLEY1969181, PhysRev.177.858}, a transition from a high-temperature cubic (space group \textit{Pm$\bar{3}$m}) to a tetragonal antiferrodistortive (AFD, space group \textit{I$4$/mcm}) phase, where \ce{TiO6} octahedra rotate around the $c$-axis with out-of-phase rotations in consecutive layers ($a^0a^0c^-$ in Glazer notation~\cite{Glazer:a09401}).

We construct the STO substrate from the fully relaxed AFD structure as a \ce{TiO2} terminated 80-atom supercell slab that has $2\times2\times4$ dimensions compared to the 5-atom cubic cell. The \textit{Pnma} SMO film also has $2\times2\times4$ dimensions compared to the 5-atom cubic cell and is \ce{MnO2} terminated. Due to lattice mismatch between STO and SMO, the SMO film experiences tensile strain of about 1.5\% when its lattice parameters are adjusted to match the STO substrate. We separate periodic images along the film normal by a 12 \AA\ vacuum and employ a dipole correction in this vacuum layer \cite{PhysRevB.59.12301}. The lowest two atomic layers of the substrate are fixed at bulk positions to mimic the presence of a large and rigid bulk.

The Brillouin zone of this thin-film system is sampled using a 6$\times$6$\times$1 Monkhorst-Pack~\cite{PhysRevB.13.5188} k-point grid. The convergence criteria for geometry relaxations were $5\cdot10^{-3}$ eV/\AA\ for forces and $10^{-5}$ eV for the total energy. The formation energy $E_{f,\mathrm{V_O}}$ of an neutral oxygen vacancy (\ce{V_O^{..}}) was calculated as described in Ref.~\onlinecite{RevModPhys.86.253}:
	\begin{equation}
    	E_{f,\ce{V_O^{..}}} = E_{\mathrm{tot}, \ce{V_O^{..}}} - E_{\mathrm{tot}, \mathrm{stoic}} + \mu_\mathrm{O},
	\end{equation}
where $\mu_\mathrm{O} = \frac{1}{2}E(\mathrm{O_2})$ is the oxygen chemical potential for which we assume the oxygen-rich limit (i.e. half the energy of an \ce{O2} molecule $E(\mathrm{O_2})$, while $E_{tot, \mathrm{V_O}}$ and  $E_{tot, stoi}$ are the total energies of the defective and stoichiometric supercells, respectively.

%%%%%%%%%%%%%%%%%%%%%%%%%%%%%%%%%%%%%%%%%%%%%%%%%%%%%%%%%%%%%%%%%%%%%%%
\section{Results and Discussion}
%%%%%%%%%%%%%%%%%%%%%%%%%%%%%%%%%%%%%%%%%%%%%%%%%%%%%%%%%%%%%%%%%%%%%%%

%%%%%%%%%%%%%%%%%%%%%%%%%%%%%%%%%%%%%%%%%%%%%%%%%%%%%%%%%%%%%%%%%%%%%%%
\subsection{Stoichiometric thin film}
%%%%%%%%%%%%%%%%%%%%%%%%%%%%%%%%%%%%%%%%%%%%%%%%%%%%%%%%%%%%%%%%%%%%%%%

\begin{figure}
  \includegraphics[width=1.0\columnwidth]{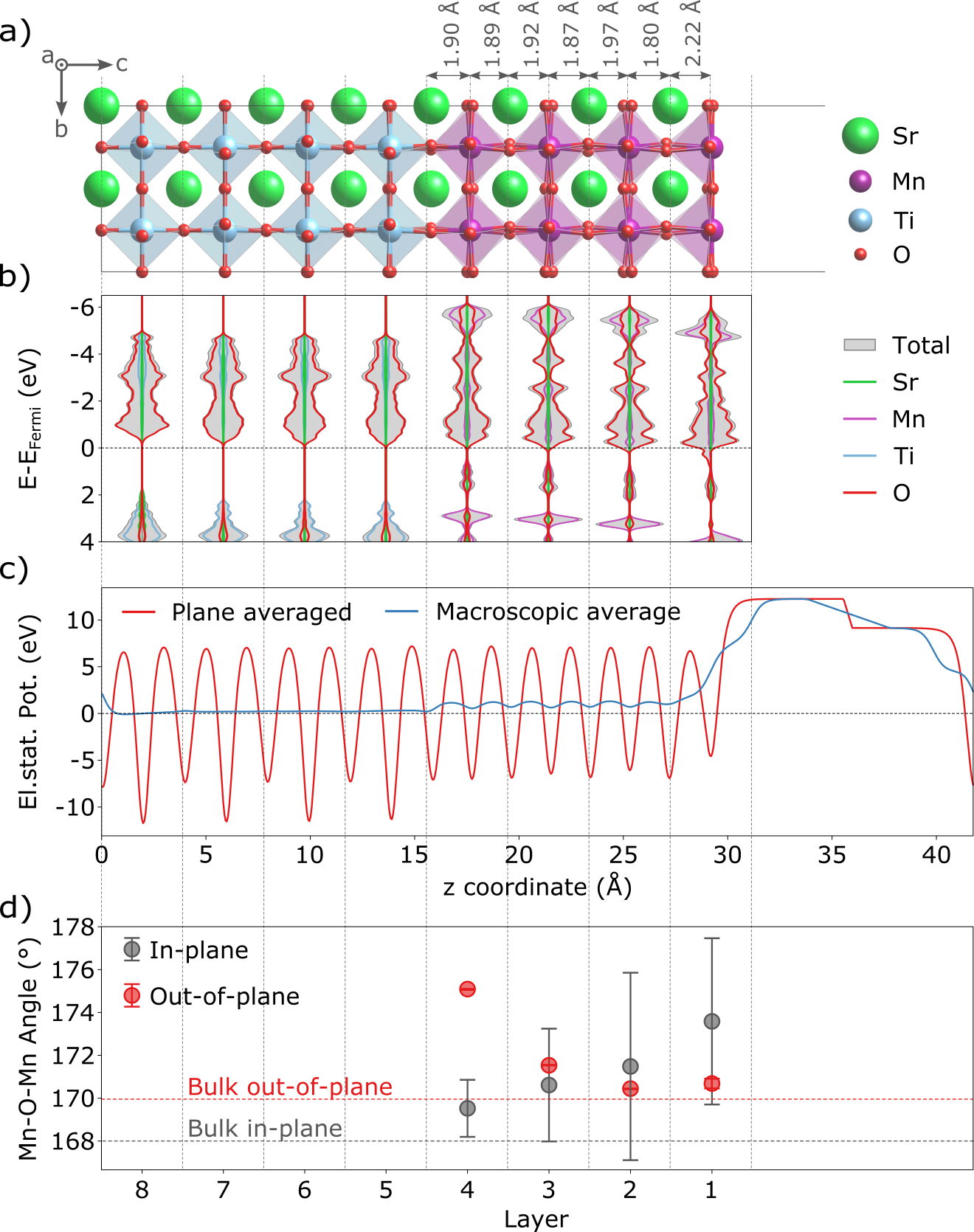}
  \caption{\label{fig:stoichiometric} Stoichiometric SMO thin film on a STO substrate: a) structure and selected interplanar spacings, b) total and projected layer-resolved density of states, c) electrostatic potential and d) Mn--O--Mn bond angles.}
\end{figure}

Our thin-film model deviates from SMO bulk in a number of ways that are expected to also affect the behavior and properties of defective films. Due to tensile strain the in-plane (IP) dimensions of SMO are expanded, while the out-of-plane (OP) dimension shrinks according to Poisson's ratio \cite{PhysRevB.94.104101}. This change in lattice parameters is accommodated by changes in Mn--O bond lengths and Mn--O--Mn bond angles (see Fig. \ref{fig:stoichiometric} a and d). In particular the IP Mn--O--Mn angles straighten out at the surface and approach 180$^\circ$, while the OP bond-angles remain close to the bulk value. We note however that the OP bond-angles are strongly affected at the interface to match the $a^0a^0c^-$ STO rotation pattern.

The truncation of Mn--O bonds at the film surface changes the crystal field of surface Mn atoms from octahedral to approximately square pyramidal. The concomitant lowering of Mn 3\textit{d} energies leads to more favorable electron accommodation at the surface compared to Mn sites in other layers. This change is visible in the layer-resolved projected density of states (Figure \ref{fig:stoichiometric} b), where at the very surface the characteristic $e_g$ peak visible around 3 eV in lower \ce{MnO2} layers is significantly destabilized and a small minority-spin peak is visible just above the Fermi energy. This change in the electronic structure leads to a ferrimagnetic surface layer (see Fig. \ref{fig:stoichiometric} b).

We evaluated the oxidation state of these surface atoms according to the method proposed in Ref. \onlinecite{Sit:2011fv} and found it to be \ce{Mn^3+}, while Mn atoms in all other layers retain their nominal \ce{Mn^4+} oxidation state. We verified that the change in oxidation state is not an artifact of the geometry. As shown in Fig. \ref{fig:SMO_pdos}, we observe the same change in electronic structure in asymmetric and symmetric SMO surface slabs, indicating that it is an intrinsic feature of \ce{MnO2}-terminated SMO surfaces and not caused by the asymmetry of the setup. Based on the projected layer-resolved densities of states (see Fig. \ref{fig:stoichiometric}b) the electrons leading to surface \ce{Mn^3+} stem from lower lying SMO layers that are slightly electron depleted.

This charge transfer towards the surface has two effects. On one hand, the reduction from Mn$^{4+}$ to Mn$^{3+}$ at the surface causes an increase in ionic radius, which manifests in an expansion of the \ce{SrO}-\ce{MnO2} interlayer spacing by about 0.3 \AA\ from 1.91 \AA\ in the bulk to 2.20 \AA\ at the surface, as visible in Fig. \ref{fig:stoichiometric}a). This structural distortion propagates into lower layers, where we observe shorter interlayer spacings below \ce{SrO} layers (from 1.91 \AA\ to 1.80 \AA\ below the first \ce{SrO} layer), while those below \ce{MnO2} layers are expanded. Such a change in geometry will affect the crystal field and could alter the excess charge accommodation upon oxygen vacancy formation. On the other hand, the presence of \ce{Mn^3+} at the surface will repel electrons from the surface. As shown in Fig. \ref{fig:stoichiometric}c), the change in electrostatic potential is greatest at the surface but a small field exists throughout the whole film. This electric field will lead to excess charges being more favorably accommodated further away from the surface. Nevertheless we note that the SMO film has empty states at lower energies than the STO substrate, which is expected to keep excess-charges in the film rather than the substrate. In the next section, we will investigate how the aforementioned changes in the ionic and electronic structure affect the formation of oxygen vacancies in the thin-film system.

%%%%%%%%%%%%%%%%%%%%%%%%%%%%%%%%%%%%%%%%%%%%%%%%%%%%%%%%%%%%%%%%%%%%%%%
\subsection{Oxygen vacancies in the thin film}
%%%%%%%%%%%%%%%%%%%%%%%%%%%%%%%%%%%%%%%%%%%%%%%%%%%%%%%%%%%%%%%%%%%%%%%

\begin{figure}
  \includegraphics[width=1.0\columnwidth]{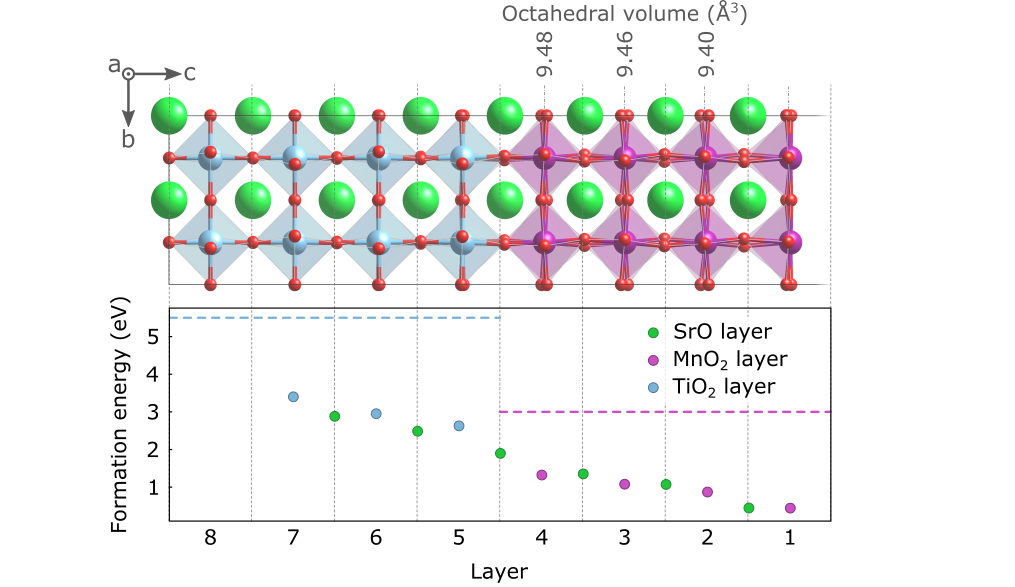}
  \caption{\label{fig:V_F}a) Formation energy for $\mathrm{V_O}$ under O rich conditions in different layers of the SMO film and the STO substrate. The dashed horizontal lines indicate the theoretical values of the formation energies in 40-atom SMO and STO bulk cells~\cite{ricca2019self, PhysRevResearch.2.023313}. \ce{MnO6} octahedral volumes are indicated above the structure.}
\end{figure}

In Fig. \ref{fig:V_F} we show the formation energy of neutral oxygen vacancies ($\mathrm{V_O}$) in the different layers of the thin film. Within each layer, multiple symmetry inequivalent $\mathrm{V_O}$ positions exist, the variation in formation energy of which is however negligible compared to the effect of the distance from the surface. We observe an increasing trend (by more than 1 eV) in formation energy from the surface to the hetero-interface, which continues into the substrate. This implies that $\mathrm{V_O}$ will have a tendency to be formed in the surface region, respectively to migrate there, if oxygen mobility is sufficiently high. It is also noteworthy that neither the formation energy in the film nor the substrate reaches the formation energy in the respective bulk phases~\cite{ricca2019self, PhysRevResearch.2.023313}, which are indicated as horizontal dashed lines in Fig. \ref{fig:V_F}. This hints at a different electronic structure of the $\mathrm{V_O}$ compared to the bulk phases.

\begin{figure}
  \includegraphics[width=1.0\columnwidth]{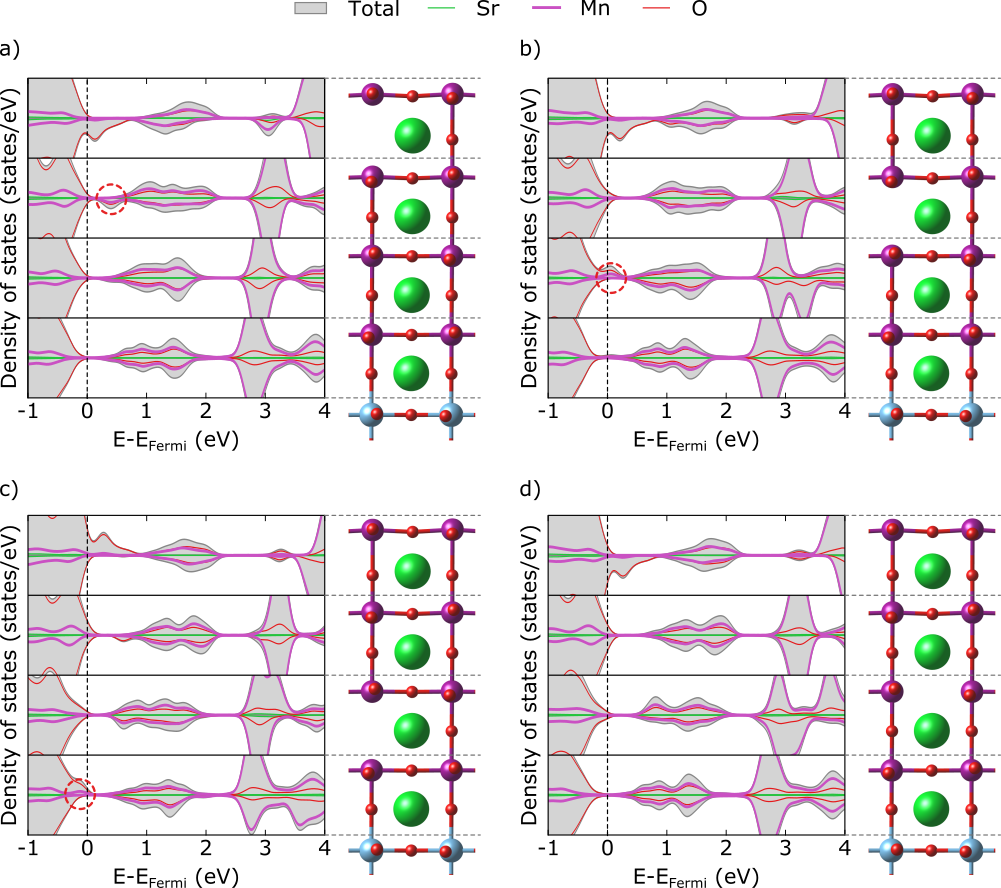}
  \caption{\label{fig:pdos} Total and projected layer-resolved density of states for $\mathrm{V_O}$ a) in the first, b) second and c) third \ce{SrO} layer and d) in the third \ce{MnO2} layer.}
\end{figure}

The metallic nature of the stoichiometric surface layer greatly complicates the analysis of the electronic structure in presence of a $\mathrm{V_O}$ and we hence start our discussion with vacancies in the \ce{SrO} layers, returning to \ce{MnO2} layers below. In bulk manganites, $\mathrm{V_O}$ formation results in a lowering of the $e_g$ states of the two \ce{Mn} atoms adjacent to the vacancy, which become the lowest unoccupied states and accomodate the excess electrons resulting from neutral $\mathrm{V_O}$ formation \cite{PhysRevB.88.054111, Becher2015, ricca2019self}. In the topmost \ce{SrO} layer of our thin film, we observe a different picture. Instead of localizing on one \ce{Mn} atom each in the surface and first subsurface \ce{MnO2} layers, we observe the typical Mn e$_g$ defect state only in the first subsurface layer (circled in Fig. \ref{fig:pdos} a). We note that this state remains unoccupied, while we observe filling of the valence-band states, notably in the second layer, that were empty due to the crystal-field changes at the stoichiometric surface. This implies that the charge primarily localizes in the layer below the vacancy, however not in the typical Mn e$_g$ defect states.

 We see the same pattern also in the second and third \ce{SrO} layer (Figs. \ref{fig:pdos} b and c), but note that with increasing depth the circled Mn e$_g$ defect state becomes occupied while the valence-band states in the first subsurface layer are again empty. We associate this with the slight upwards band bending at the surface induced by the \ce{Mn^3+} ions that causes the defect state in layers further from the surface to be lower in energy than closer to the surface. This is also in agreement with the fact that electrons always localize in layers below the defect, i.e. further away from the surface. Moreover, the reduction-less excess-charge accommodation in valence-band states explains the lower formation energy close to the surface, which gradually increases due to increased defect-state occupation in layers further from the surface.

The only exception to this excess-charge accommodation pattern occurs in the \ce{SrO} layer at the SMO-STO interface, where the $\mathrm{V_O}$ is formed between a \ce{Mn} and a \ce{Ti} atom. Here the reduction occurs on the \ce{Mn} above the $\mathrm{V_O}$, which can be rationalized by the lower lying empty \ce{Mn} states compared to the empty \ce{Ti} states (see Fig. \ref{fig:stoichiometric} b). This change in electronic structure can account for the marked increase in formation energy for this layer, compared to layers closer to the surface (see Fig. \ref{fig:V_F}).

For $\mathrm{V_O}$ in the \ce{MnO2} layers (exemplified for the third \ce{MnO2} layer in Fig. \ref{fig:pdos} d) we observe an asymmetry of e$_g$ peaks in the majority and minority spin channels. Indeed, we observe significantly more Mn e$_g$ density of states in the valence band of the vacancy layer than for the other layers. We thus conclude that one of the two Mn sites next to the $\mathrm{V_O}$ gets reduced (the spin-down Mn in the G-AFM order in this case), whereas the other (the spin-up Mn) does not get reduced, the remaining electron filling the empty valence-band states.

The variation in bond lengths discussed for the stoichiometric surface leads to changes in the octahedral volumes and one could, via chemical expansion arguments~\cite{Adler2001chemical, PhysRevB.88.054111}, expect the $\mathrm{V_O}$ formation energy to be inversely correlated with the polyhedral volume in the stoichiometric structure. While the strongly expanded truncated polyhedron at the surface is indeed associated with a low formation energy, this trend does not continue towards the interface (see octahedral volumes Fig. \ref{fig:V_F}) since the octahedral volumes increase as does the formation energy. We therefore believe the $\mathrm{V_O}$ formation energy to be dominated by the band bending induced by the reduced surface \ce{Mn^3+} ions.

%%%%%%%%%%%%%%%%%%%%%%%%%%%%%%%%%%%%%%%%%%%%%%%%%%%%%%%%%%%%%%%%%%%%%%%
\subsection{Oxygen vacancies in the substrate}
%%%%%%%%%%%%%%%%%%%%%%%%%%%%%%%%%%%%%%%%%%%%%%%%%%%%%%%%%%%%%%%%%%%%%%%

\begin{figure}
  \includegraphics[width=1.0 \columnwidth]{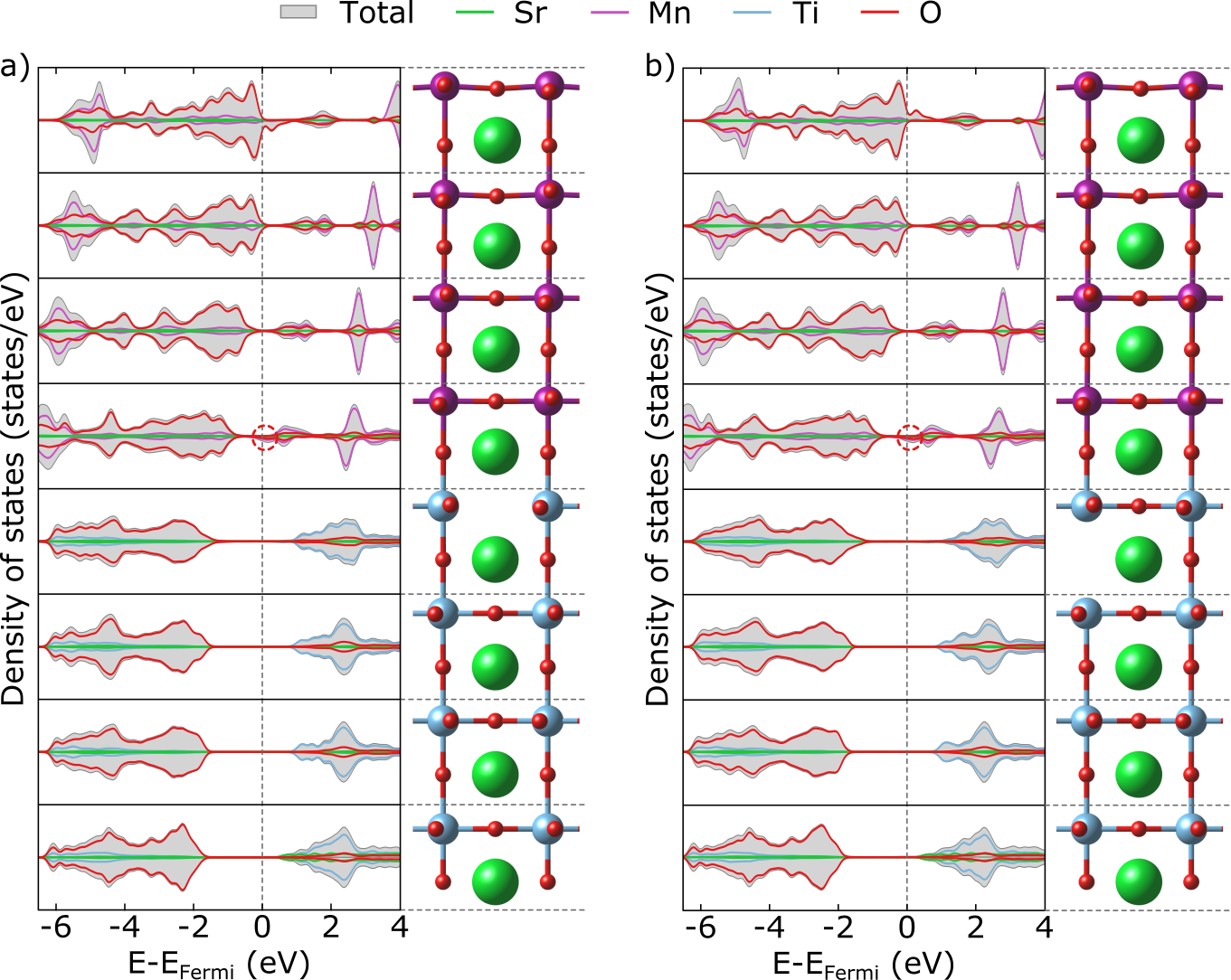}
  \caption{\label{fig:pdos_STO} Total and projected layer-resolved density of states for $\mathrm{V_O}$ a) in the first \ce{TiO2} layer and b) in the first \ce{SrO} layer of the STO substrate.}
\end{figure}

In bulk STO, the excess electrons associated with a neutral $\mathrm{V_O}$ localize in F-center like states derived from Ti $e_{g}$ orbitals and possibly also populate the conduction band if a spin triplet state is allowed \cite{Hou2010, Lin2012, Mitra2012, Choi2013, PhysRevResearch.2.023313}. In our thin film setup we observe, on the contrary, no localization of electrons in the vicinity of a $\mathrm{V_O}$ created in the STO substrate. On the contrary both for a $\mathrm{V_O}$ created in a \ce{TiO2} or in a \ce{SrO} layer of the substrate, the excess electrons primarily localize on Mn states at the bottom of the SMO thin film but also fill the above-mentioned valence-band states induced by surface \ce{Mn^3+} (see Fig.~\ref{fig:pdos_STO}).

This separation of the $\mathrm{V_O}$ defect from it's excess charge occurs - at least within the scale of our computational model - independently of the distance of the $\mathrm{V_O}$ from the interface and can be rationalized by the significant energy difference between empty Mn and Ti states (see Fig. \ref{fig:stoichiometric}). The defect formation energy can be significantly lowered by accommodating electrons in energetically more favorable \ce{Mn} states than in \ce{Ti} states that are either in the STO conduction band or just below the conduction band edge.

This interfacial charge transfer is the reason behind the reduced formation energy compared to bulk STO~\cite{PhysRevResearch.2.023313}. Nevertheless we note that $\mathrm{V_O}$ formation energies will always be significantly larger in the substrate compared to the thin film and that $\mathrm{V_O}$ will have a driving force for migration from the STO substrate into the SMO thin film.

%%%%%%%%%%%%%%%%%%%%%%%%%%%%%%%%%%%%%%%%%%%%%%%%%%%%%%%%%%%%%%%%%%%%%%%
\section{Conclusions}
%%%%%%%%%%%%%%%%%%%%%%%%%%%%%%%%%%%%%%%%%%%%%%%%%%%%%%%%%%%%%%%%%%%%%%%

In this work, we investigated the effect of the surface and hetero-interface on oxygen vacancies in \ce{SrMnO3} thin films grown on a \ce{SrTiO3} substrate. We show that the altered crystal field at a \ce{MnO2} terminated film surface leads to a charge transfer from valence-band states in lower layers to the surface, forming reduced \ce{Mn^3+} ions at the very surface. This alteration affects oxygen-vacancy formation in the film in two ways. On one hand the holes created in the film due to the above-mentioned charge transfer accomodate some of the excess charge induced by neutral oxygen vacancies. On the other hand, the remaining excess charge that localizes in the vicinity of the vacancy is affected by the surface-\ce{Mn^3+} induced band bending, which leads to formation energies that gradually increase from the surface to the interface. We further show that vacancies in the \ce{SrTiO3} substrate have larger formation energies compared to the \ce{SrMnO3} film but transfer electrons to the film, which lowers their formation energy compared to bulk \ce{SrTiO3}.

A previous experimental study~\cite{Wang2020} on the same thin film system, used EELS to detect an increased electron density in the \ce{MnO2} layer closest to the substrate. This was explained based on multiple origins, among them oxygen vacancy formation in the bulk and a reduction of the substrate under deposition conditions. Based on our findings, oxygen vacancies in the \ce{SrMnO3} film would have to reside at the very interface to induce charges in that layer. Since there is no driving force for vacancy accumulation close to the interface and since vacancies in other layers would also affect the Mn oxidation state in these layers, oxygen vacancies in \ce{SrMnO3} seem an unlikely source for the experimental observations. On the other hand, oxygen vacancies in the substrate (likely to be formed under vacuum deposition conditions) would consistently lead to charges in the layer observed by experiment and are a more likely scenario.

Our results show that oxygen vacancy formation in a thin-film geometry can be significantly different from a bulk or strained bulk situation. Not only can the film surface induce subtle changes in the electronic structure that affect excess charge accommodation and stability of oxygen vacancies, but the film can also attract excess electrons from the substrate. These effects are not captured in established strained bulk calculations and are expected to strongly depend on the properties of the film and the substrate. As such a fundamental understanding of defects in oxide thin films has to involve surface and interface effects in addition to strain effects.

%%%%%%%%%%%%%%%%%%%%%%%%%%%%%%%%%%%%%%%%%%%%%%%%%%%%%%%%%%%%%%%%%%%%%%%
\section{Acknowledgements}
%%%%%%%%%%%%%%%%%%%%%%%%%%%%%%%%%%%%%%%%%%%%%%%%%%%%%%%%%%%%%%%%%%%%%%%

This work was supported by the Swiss National Science Foundation under Project 200021\_178791. Calculations were performed on UBELIX (http://www.id.unibe.ch/hpc), the HPC cluster at the University of Bern and on Piz Daint at the Swiss Supercomputing Center CSCS under projects s955 and s1033.

\bibliography{references}

%%%%%%%%%%%%%%%%%%%%%%%%%%%%%%%%%%%%%%%%%%%%%%%%%%%%%%%%%%%%%%%%%%%
%%%                        SUPPLEMENTARY                        %%%
%%%%%%%%%%%%%%%%%%%%%%%%%%%%%%%%%%%%%%%%%%%%%%%%%%%%%%%%%%%%%%%%%%%

%reset all style and numbering
\clearpage
\clearpage %needed for two-page reference section
\setcounter{page}{1}
\renewcommand{\thetable}{S\arabic{table}}  
\setcounter{table}{0}
\renewcommand{\thefigure}{S\arabic{figure}}
\setcounter{figure}{0}
\renewcommand{\thesection}{S\arabic{section}}
\setcounter{section}{0}
\renewcommand{\theequation}{S\arabic{equation}}
\setcounter{equation}{0}
\onecolumngrid

%create title
\begin{center}
\textbf{Supplementary information for\\\vspace{0.5 cm}}
\textbf{\large Surface and interface effects in oxygen deficient \ce{SrMnO3} thin films grown on \ce{SrTiO3}\\\vspace{0.3 cm}}
Moloud Kaviani and Ulrich Aschauer

\small
\textit{Department of Chemistry, Biochemistry and Pharmaceutical Sciences, University of Bern, Freiestrasse 3, CH-3012 Bern, Switzerland}

(Dated: \today)
\end{center}

%%%%%%%%%%%%%%%%%%%%%%%%%%%%%%%%%%%%%%%%%%%%%%%%%%%%%%%%%%%%%%%%%%%%%%%
\section{Effect of film symmetry}
%%%%%%%%%%%%%%%%%%%%%%%%%%%%%%%%%%%%%%%%%%%%%%%%%%%%%%%%%%%%%%%%%%%%%%%

\begin{figure}[h]
  \includegraphics[width=0.75\columnwidth]{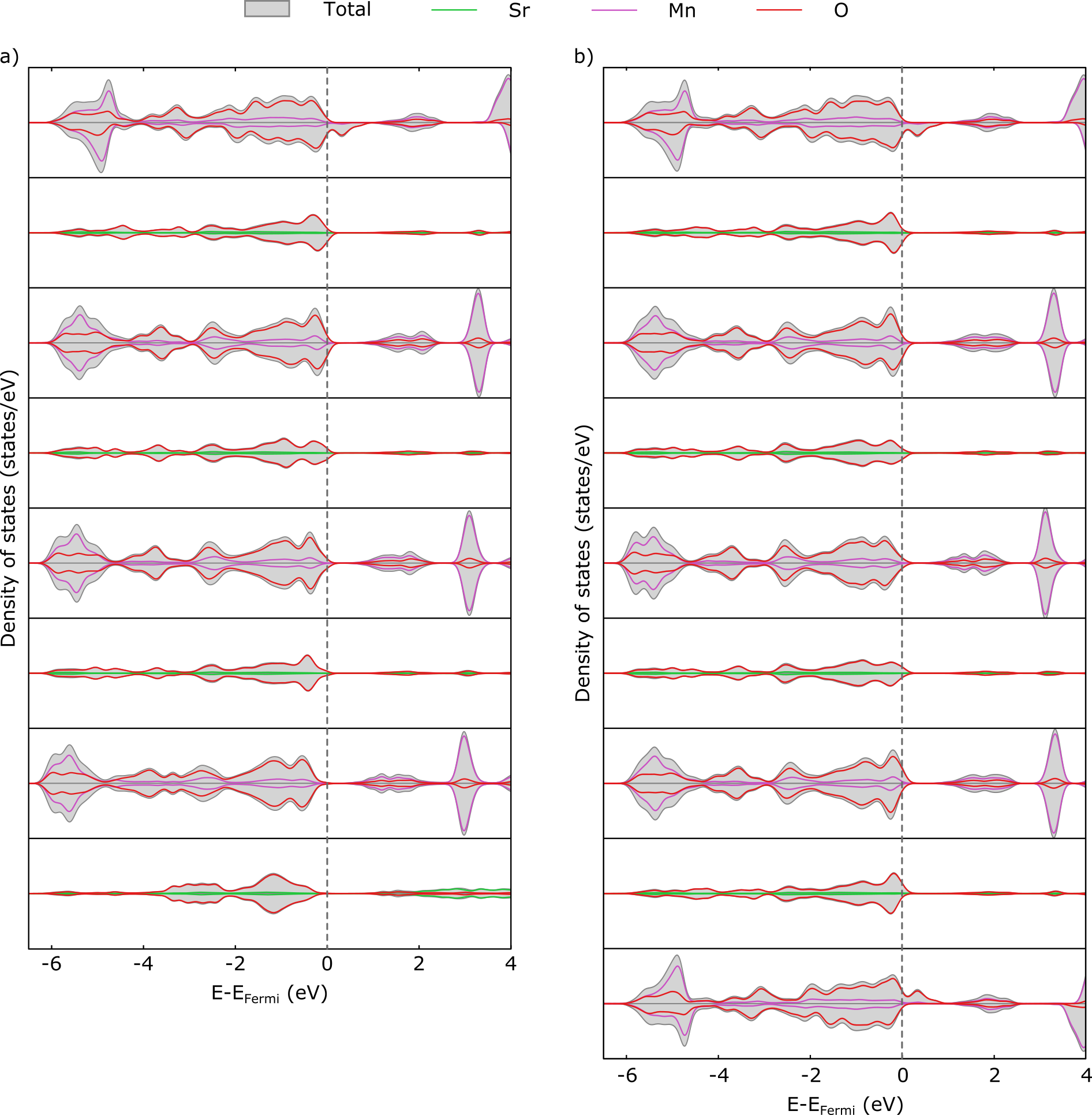}
  \caption{\label{fig:SMO_pdos} Total and projected layer-resolved density of states for a) the stoichiometric, asymmetric \ce{SrMnO3} slab and b) the non-stoichiometric, symmetric \ce{SrMnO3} slab.}
\end{figure}

In Fig. \ref{fig:SMO_pdos}, we show the layer-resolved projected density of states for an asymmetric \ce{SrMnO3} slab with a \ce{MnO2} top surface and a \ce{SrO} bottom surface as well as for a symmetric slab with two \ce{MnO2 surfaces}. The key result is that when both surfaces are \ce{MnO2}-terminated and allowed to relax, they develop the same electronic structure as the top surface of the asymmetric slab. Since in the symmetric slab, no spurious electric field due to the presence of two different surface terminations is present, we associate the changed surface electronic structure with crystal-field changes from octahedral to square pyramidal, as mentioned in the main text, rather than it being an artifact associated with the asymmetric geometry of the film.

\end{document}